# Processing issues in SiC and GaN power devices technology: the cases of 4H-SiC planar MOSFET and recessed hybrid GaN MISHEMT

F. Roccaforte *, G. Greco, P. Fiorenza
* Consiglio Nazionale delle Ricerche – Istituto per la Microelettronica e Microsistemi (CNR-IMM),
Strada VIII, n. 5 – Zona Industriale, 95121 Catania - Italy
E-mail: **fabrizio.roccaforte@imm.cnr.it**

*Abstract*—*This paper aims to give a short overview on some relevant processing issues existing in SiC and GaN power devices technology. The main focus is put on the importance of the channel mobility in transistors, which is one of the keys to reduce $R_{ON}$ and power dissipation. Specifically, in the case of the 4H-SiC planar MOSFETs the most common solutions and recent trends to improve the channel mobility are presented. In the case of GaN, the viable routes to achieve normally-off HEMTs operation are briefly introduced, giving emphasis to the case of the recessed hybrid MISHEMT.*

## 1. Introduction

The worldwide increasing need of electric energy is a serious concern in our society. In fact, the energy consumption in the world is estimated to increase of 40% in the next two decades [1] and the largest fraction (up to 60%) of the consumed energy will be electric energy. Hence, energy efficiency has become a challenge in modern semiconductor power devices technologies, to ultimately reduce the global energy consumption.

Currently, power electronics market is almost entirely based on Silicon (Si) devices [2]. However, Si-based power electronics has reached its performance limits, in terms of maximum power levels, frequency and operation temperatures. Hence, the only way to overcome the physical limits of Si is a radical innovation of the technology for discrete semiconductor power devices.

In this context, due to their excellent physical properties [3], the most popular wide band gap (WBG) semiconductors, silicon carbide (4H-SiC) and gallium nitride (GaN), are considered the best materials to replace Si in the future high efficient power electronics. **Fig. 1** shows a graphical comparison of some relevant physical properties of Si, SiC and GaN. As can be seen, the large values of energy gap and critical electric field allow these materials to operate at high breakdown voltages ($B_V$). The high saturated electron velocity enables superior performances under high frequency operation. Finally, the high thermal conductivity (in the case of SiC) is an important feature that guarantees an easy heat dissipation for operation at high temperature and high current levels.

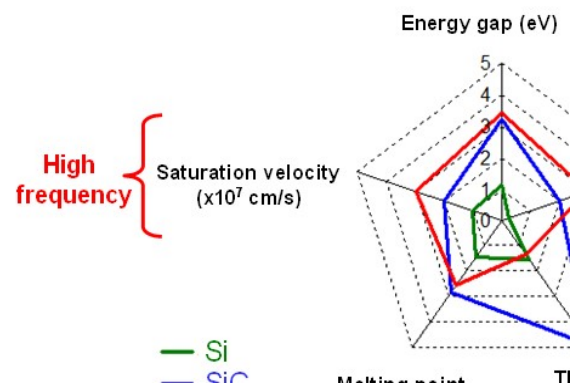

**Fig. 1** Comparison of Si, SiC and GaN relevant properties for power devices applications.

These outstanding properties of SiC and GaN enable to design transistors with a smaller ON-resistance ($R_{ON}$) and smaller parasitic capacitances with respect to the Si counterparts for a fixed targeted maximum operation voltage. The direct impact of a lower $R_{ON}$ is a reduction of the total power dissipation [4]. Hence, SiC and GaN devices can find several applications in power electronics in many important fields. To visualize the huge potential of these materials, ***Fig. 2*** depicts the major applications of WBG power devices in a power versus voltage chart. As can be seen the possible application areas enter our daily life, e.g., consumer electronics (PFC/power supply, audio amplifiers,…), EV/HEV automotive components (converters, battery chargers, ….), industrial applications (motor drives,..), renewable energies (PV-inverters,…), transportations, etc.

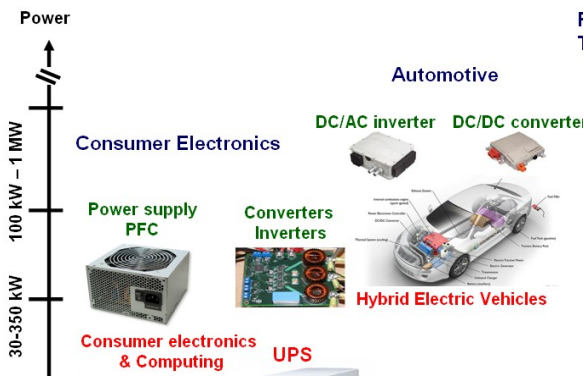

**Fig. 2** Main application areas of SiC and GaN power devices.

Today, while several 4H-SiC and GaN transistors with excellent performances have already reached the market, there are still some important physical problems related to the fabrication processes of these devices, which are still object of intensive investigation by the scientific community.

This paper aims to give a brief overview on some current processing issues encountered in SiC and GaN power devices, with a focus on transistors technology. In particular, the most common approaches to improve the MOS interface quality in 4H-SiC planar MOSFETs are presented, highlighting their advantages and limitations. Moreover, the feasible solutions to achieve normally-off operation in GaN HEMTs are presented, with special attention to the case of the recessed hybrid MISHEMT.

.

## 2. 4H-SiC MOSFET

One of the long standing problems in 4H-SiC planar MOSFETs technology is the low inversion channel mobility, especially below 1 kV, i.e., where the channel mobility can represent an important contribution to the total $R_{ON}$. This latter can be clearly seen in ***Fig. 3***, reporting the specific $R_{ON}$ as a function of the breakdown voltage $B_V$ for different values of the inversion layer channel mobility [5].

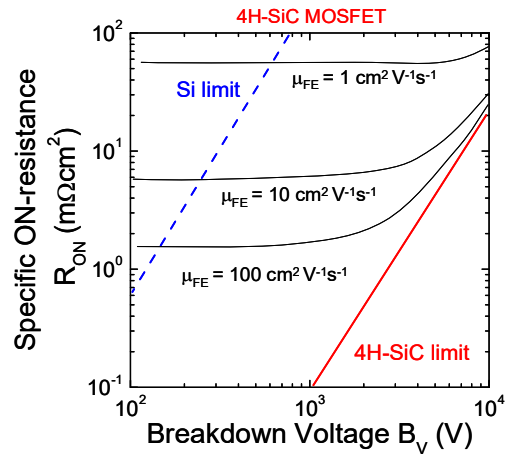

**Fig. 3** Specific ON-resistance $R_{ON}$ versus breakdown voltage $B_V$ for 4H-SiC MOSFETs, estimated for different values of the channel mobility $\mu_{FE}$.

The problem of the channel mobility in 4H-SiC MOSFETs has been recently reviewed by *Cabello et al.* [6]. In general, low values of the channel mobility (typically < 5-10 cm$^2$V$^{-1}$s$^{-1}$) are obtained with thermal SiO$_2$ gates, due to the high density of interface traps ($D_{it}$) near the conduction band edge [7,8], determining Coulombic scattering effects by charges trapped at the interface states and inside the oxide [9,10]. Hence, post deposition annealing or innovative gate oxide processes are mandatory to increase the channel mobility and decrease the $R_{ON}$.

***Fig. 4*** reports the values of the field effect mobility $\mu_{FE}$ of 4H-SiC planar MOSFETs for different treatments of the gate oxide. For a direct comparison of the data, the mobility curves are reported as a function of the difference between the gate voltage and the threshold voltage ($V_g$-$V_{th}$). The mobility curve $\mu_{FE}$ of an "untreated" dry oxide is also reported as a reference ( < 5 cm$^2$V$^{-1}$s$^{-1}$). As can be seen, the experimental $\mu_{FE}$ mobility curves versus gate voltage typically exhibit a maximum (peak mobility) after the threshold voltage $V_{th}$ is reached. Then, the channel mobility slightly decreases with increasing the gate voltage (i.e., with increasing the transversal electric field) due to the dominance of phonon and interface scattering mechanisms [11].

To improve the channel mobility, nitridation processes of the gate oxides, i.e., post-oxidation-annealing (POA) or post-deposition-annealing (PDA) in nitrogen-rich atmospheres (NO or $N_2O$) in the temperature range 1000-1300 °C, have been introduced at the end of the 90's [12,13,14,15].

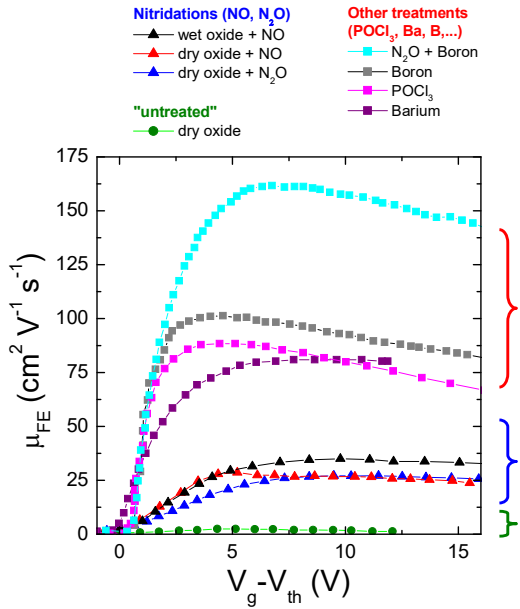

**Fig. 4** Field effect mobility ($\mu_{FE}$) as a function of the difference between the gate voltage and the threshold voltage ($V_g$-$V_{th}$) in 4H-SiC planar MOSFETs fabricated employing different gate oxides treatments. The data are from Ref. [2] and references therein.

The improvement of the mobility (up to 25-50 $cm^2V^{-1}s^{-1}$) obtained upon nitridation is typically accompanied by a reduction of the interface state density $D_{it}$ down to the low $10^{12}$ $eV^{-1}cm^{-2}$ range.

As an alternative to the nitridations, the introduction of different species in the gate oxide has been considered to passivate the $SiO_2$/SiC interface states and increase the mobility $\mu_{FE}$. *Okamoto et al.* [16] demonstrated that an annealing of the gate oxide in phosphoryl chloride ($POCl_3$) can significantly increase the 4H-SiC MOSFETs mobility (89 $cm^2V^{-1}s^{-1}$). Later, other authors explored similar phosphorous-based passivation routes, obtaining mobility values higher than 100 $cm^2V^{-1}s^{-1}$, with $D_{it}$ in the $10^{11}$ $cm^{-2}eV^{-1}$ range [17,18,19].

During nitridation ($N_2O$ or NO) or $POCl_3$ processes, the presence of n-type dopant (i.e., nitrogen and phosphorous) in the annealing atmosphere determines notable electrical changes in the $SiO_2$/SiC interface. In fact, nitrogen and phosphorous atoms can be incorporated in the SiC substrate during annealing, and act as n-type shallow donors in the material [20,21]. Using scanning probe microscopy analyses at the $SiO_2$/SiC interface allowed to demonstrate the "counter doping effect" of the p-type implanted regions in the MOSFET channel [17,22]. These measurements also showed a higher electrically active phosphorous incorporation in $POCl_3$ with respect to the active nitrogen incorporated in $N_2O$ [22].

In spite of the high channel mobility, the drawback of the $POCl_3$ annealing is the poor reliability of the gate oxides, caused by the large amount of charge traps in the $SiO_2$ network after a phosphorous incorporation [23]. Some research groups proposed other phosphorous-based processes ($POCl_3$ pre-annealings before oxide deposition, combination of N- and P-based annealings, P-ion-implantation), with promising results in terms of mobility and improvement of the $V_{th}$ stability [18,19,24,25,26]. More recently, channel mobility values > 100 $cm^2V^{-1}s^{-1}$ have been obtained using other group-V elements (e.g., As, Sb), in conjunction with nitric oxide (NO) post-oxidation annealing [27]. However, the $\mu_{FE}$ curves of As- or Sb-doped 4H-MOSFETs channels exhibit pronounced maxima at low electric fields, but decrease rapidly at high fields (e.g.>10 V). Hence, As- or Sb-counter-doping appears of limited effectiveness in real devices [6].

Another recent approach to increase the

4H-SiC MOSFET mobility is the use of Boron (B). *Okamoto et al.* [28] achieved a mobility of about 100 cm$^2$V$^{-1}$s$^{-1}$ using Boron thermal diffusion (by a planar BN diffusion source) into a dry oxide. Since B is an acceptor for SiC, "counter doping" does not occur and cannot explain the increased mobility. Hence, these results were attributed to a stress relaxation of the interface by the incorporation of B-atoms in the SiO$_2$ matrix [29]. This process was recently optimized, by combining the N$_2$O oxinitridation with B-diffusion [30,31]. In this way, a peak mobility of 160 cm$^2$V$^{-1}$s$^{-1}$ has been obtained, while a stable threshold voltage $V_{th}$ at least at room temperature [6].

Finally, the use of alkali or alkaline earth elements (Rb, Cs, Sr, Ba,…) has been proposed to passivate the SiO$_2$/4H-SiC interface states and increase the 4H-SiC MOSFET mobility. These processes typically consist in the deposition of a thin layer of alkali/alkaline-earth material on SiC, followed by the deposition and post-annealing (in O$_2$ or O$_2$/N$_2$ ambient)) of SiO$_2$ gate oxide. Among various elements the most promising results were achieved with Sr and Ba, with mobility values of $\mu_{FE}$ up to 65 and 110 cm$^2$V$^{-1}$s$^{-1}$, respectively [32,33,34]. It has been also shown that Ba incorporation allows to obtain a threshold voltage stability under stress at 175 °C and 2 MV/cm gate bias. The beneficial role of Ba was explained in term of interface stress release using transmission electron microscopy analysis. In particular, the tensile strain of the SiC region close to the SiO$_2$/SiC interface is released in the presence of an oxidized Ba interlayer. Such an "unstrained" interface is the key factor for the increase of the channel mobility [35,36].

Despite the significant improvements of the channel mobility achievable with the aforementioned approaches, most of these processes are still far to be employed in "real" devices, since they are affected by threshold voltage $V_{th}$ instability issues. Hence, nitridation of the gate oxide remains the process of choice in the fabrication of state-of-the-art 4H-SiC MOSFETs.

## 3. Normally-OFF GaN HEMTs

In principle, due to its higher critical electric field (***Fig. 1***) one may expect from GaN a better high voltage operation behavior than SiC. However, a large density of defects is still present in GaN-based materials, which hinders to reach the theoretical electric field strength. Moreover, the lack of high quality large diameter bulk GaN substrates does not allow the realization of power devices with vertical architectures, as needed for a high breakdown voltages at low $R_{ON}$. Consequently, lateral heterojunction devices are nowadays the preferred solution to fabricate GaN-based transistors. In particular, GaN high electron mobility transistors (HEMTs) are normally-ON devices, due to the presence of the two dimensional electron gas (2DEG) in AlGaN/GaN heterostructures. However, power electronics applications typically require normally-OFF devices, to guarantee fail-safe operation and gate drivers simplicity [37,38,39]. Hence, significant efforts have been devoted in the last decade to develop physical methods to control the 2DEG in the channel and obtain HEMT with a positive threshold voltage $V_{th}$.

The use of a p-GaN gate is currently the only commercial solution for normally-OFF GaN HEMTs [40]. *Greco et al.* [41] recently summarized in a review the most relevant processing issues in normally-OFF HEMTs with the p-GaN gate approach. Hence, this layout will be not subject of discussion in the present paper.

Another promising approach consists in the complete removal of the AlGaN barrier under the gate [42,43], creating a metal insulator semiconductor (MIS) recessed-gate hybrid HEMT (MISHEMT). The recessed-gate hybrid MISHEMT enables to have a positive threshold voltage $V_{th}$ of the MIS channel, preserving a low on resistance $R_{ON}$ in the access regions. The most important part of such a device is the recessed channel, in which the carriers mobility is influenced by several factors (roughness of the etched surface, defects, quality of the gate insulator, etc). Hence, characterizing the properties of insulator/GaN interface and understanding the mechanisms limiting the channel mobility are key aspects for the progress of the recessed-gate MISHEMTs technology.

Various dielectric materials have been proposed to fabricate recessed-gate normally-OFF hybrid GaN MISHEMTs (SiO$_2$, SiN, Al$_2$O$_3$, AlN/SiN....) [44,45]. As in the case of standard MOSFET, the field effect mobility $\mu_{FE}$ is an important parameter that must be optimized in order to reduce the total device $R_{ON}$ [2].

Similarly to the case of a MOSFET, also in the MISHEMT the field effect mobility $\mu_{FE}$ increases with the gate bias $V_g$ up to a maximum $\mu_{FE(peak)}$ and then decreases at high electric fields.

As can be seen in **Table 1** the values of peak mobility $\mu_{FE(peak)}$ reported in literature vary approximately in the range 30–250 cm$^2$V$^1$s$^{-1}$, with threshold voltage values V$_{th}$ of 1-2Volts. The specific on-resistance $R_{ON}$ (taken at gate bias values of V$_g$> 15V) lies in the interval 7–20 Ωmm.

**Table 1.** Values of mobility $\mu_{FE}$ and threshold voltage $V_{th}$ reported for normally-OFF recessed hybrid GaN MISHEMTs, employing different gate insulators.

| Gate insulator and thickness | $\mu_{FE(peak)}$ (cm$^2$V$^{-1}$s$^{-1}$) | V$_{th}$ (V) | Ref. |
|---|---|---|---|
| SiN $_{(20nm)}$ | 120 | 5.2 | [42] |
| Al$_2$O$_3$ $_{(30nm)}$ | 225 | 2 | [46] |
| Al$_2$O$_3$ $_{(38nm)}$ | 55 | 3.5 | [47] |
| SiO$_2$ $_{(60nm)}$ | 166 | 3.7 | [48] |
| SiO$_2$ $_{(60nm)}$ | 94 | 2.4 | [48] |
| Al$_2$O$_3$ $_{(10nm)}$ | 251 | 1.7 | [49] |
| Al$_2$O$_3$ $_{(20nm)}$ | 148 | 2.9 | [50] |
| Al$_2$O$_3$ $_{(30nm)}$ | 170 | 3.5 | [51] |
| SiN$_{(2nm)LT}$/ SiN$_{(15nm)HT}$ | 160 | 2.37 | [52] |
| SiN $_{(17nm)\ HT}$ | 38 | 1.28 | [52] |
| SiN $_{(20nm)}$ | 203 | 1.2 | [53] |
| Al$_2$O$_3$ $_{(18nm)}$ | 65 | 7.6 | [54] |
| SiO$_2$ $_{(50nm)}$ | 110 | 0.7 | [55] |
| AlN$_{(7nm)}$/ SiN $_{(7nm)}$ | 180 | 1.2 | [56] |
| Al$_2$O$_3$ $_{(5nm)}$/SiN$_{(25nm)}$ | 122 | 1.7 | [57] |

From these data, it is not simple to find a correlation between the values of $\mu_{FE(peak)}$ and $R_{ON}$, due to the fact that the reported devices are extremely different (in terms of geometry, recession processes to prepare the channel region, etc.).

However, besides its maximum, it is important to have high channel mobility values also at the operative electric field.

In order to predict the device behavior under operative conditions, it is very important to understand the dependence of the mobility on different parameters (surface roughness, interface traps, electric field, temperature, etc.).

*Fiorenza et al.* [55] investigated the temperature and field dependence of the channel mobility in recessed-gate hybrid GaN MISHEMTs using SiO$_2$ as gate insulator. **Fig. 5** reports the peak mobility $\mu_{FE\ (peak)}$ (the maxima of the $\mu_{FE}$ curves) as a function of the temperature for a recessed SiO$_2$/GaN MISHEMT [55]. From this figure, it is possible to see that the experimental $\mu_{FE\ (peak)}$ data slightly decrease with increasing the measurement temperature. Assuming a formalism analogous to a standard MOSFET, the channel mobility was expressed including in the Matthiessen's rule different scattering contributions, i.e., the bulk mobility factor ($\mu_B$), the acoustic-phonon scattering ($\mu_{AC}$), the surface roughness scattering ($\mu_{SR}$), and the Coulomb scattering ($\mu_C$) due to interface charges [55].

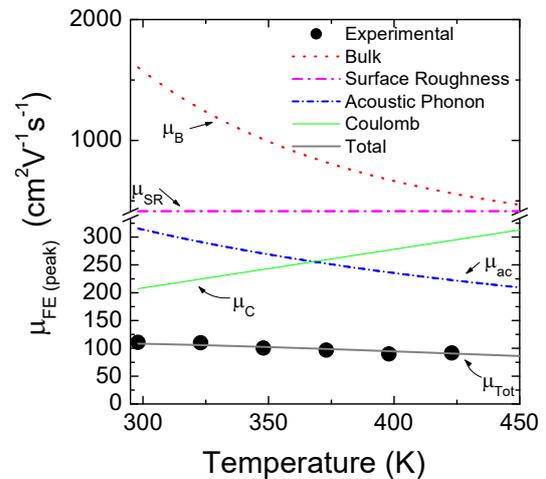

**Fig. 5.** Peak mobility values μ$_{FE\,(peak)}$ as a function of the temperature for a recessed hybrid MISHEMT using SiO$_2$ as gate insulator. The experimental data were fitted with a total mobility curve (μ$_{TOT}$) including the different contributions in the Matthiessen's rule (μ$_B$, μ$_{SR}$, μ$_{AC}$, and μ$_C$). The data are from Ref. [55].

The single contributions to the total mobility depend on several physical features of the insulator/GaN interface (roughness, doping, interface traps, etc.). Some of these parameters can be determined by direct electrical and morphological analyses of the channel region [55]. In particular, using the experimental values of interface trapped charges ($Q_{trap} = 1.35 \times 10^{12}$ cm$^{-2}$) and surface roughness (RMS = 0.15 nm), determined by C-V and AFM measurements respectively, it was possible to extract the single contributions to the mobility. The total mobility μ$_{TOT}$ and the single contributions are also reported in **Fig. 5**, and show a good agreement with the experimental data.

The temperature dependence of the peak mobility suggests that the main limiting factors to the carrier flow in the channel are the surface roughness (μ$_{SR}$), the acoustic phonon (μ$_{AC}$), and the Coulomb scattering (μ$_C$) contributions.

Hence, the optimization of the insulator/GaN interface in the recessed channel in terms of roughness and the interface trap density is a fundamental issue to improve the mobility.

In this context, the structural and electronic quality of the recessed interface could be improved by using an innovative AlN/SiN stack, grown by metal organic chemical vapour deposition (MOCVD), as gate insulating material [56]. In particular, in this case the overall the interface states $D_{it}$ was reduced with respect to the SiO$_2$/GaN MISHEMT, as can be seen in the $D_{it}$ versus energy plot in **Fig. 6a**. In fact, the total amount of trapped charge in AlN/SiN ($Q_{trap} = 6.4 \times 10^{11}$ cm$^{-2}$), i.e., the integral of the interface state density over the energy, is less than one half of the value obtained in SiO$_2$ ($1.35 \times 10^{12}$ cm$^{-2}$). This improvement allows the increase of the peak mobility from 110 cm$^2$V$^{-1}$s$^{-1}$ (SiO$_2$ gate dielectric) up to 180 cm$^2$V$^{-1}$s$^{-1}$ (AlN/SiN gate dielectric), shown in **Fig. 6b**.

The high on/off ratio observed in the case of the transistors employing AlN/SiN makes this system very promising for power switching applications [56].

As conclusive remark, it must be mentioned that channel mobility μ$_{FE}$ and the ON-resistance $R_{ON}$ are not the only parameters to be considered in this technology. In fact, recessed-gate hybrid GaN MISHEMTs are often affected by instability phenomena of the threshold voltage $V_{th}$, when subjected to gate bias stresses.

These effects are associated to the charge trapping/de-trapping of defects located at the insulator/GaN interface and/or in the bulk of the gate insulator [58].

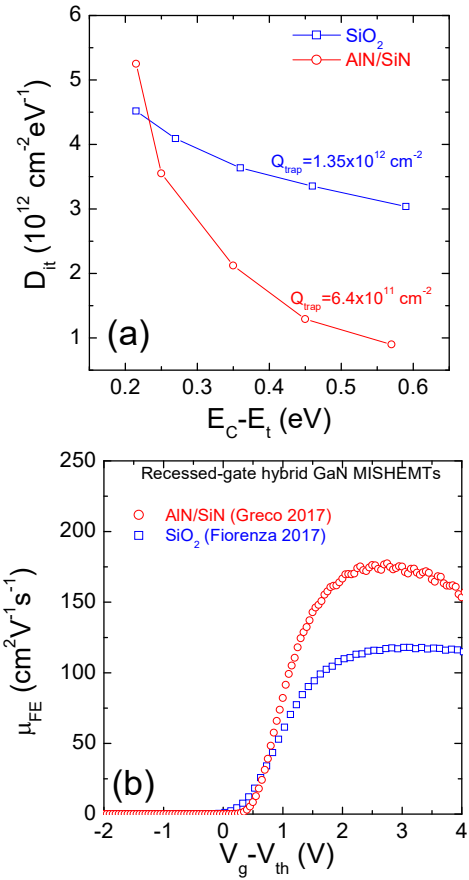

**Fig. 6.** (a) Interface state density $D_{it}$ measured both in the SiO$_2$/GaN and AlN/SiN/GaN gates in hybrid MISHEMTs. (b) Channel mobility μ$_{FE}$ as a function of the difference between the gate voltage and the threshold voltage ($V_g$-$V_{th}$) for recessed-gate hybrid GaN MISHEMTs using SiO$_2$, AlN/SiN and Al$_2$O$_3$ as gate insulators. The data are from Refs. [55,56].

Hence, a careful optimization of the properties of the interface and of the insulating materials is the route towards the achievement of stable MISHEMT devices.

## 4. Summary

In this paper, a short summary of some processing issues in SiC and GaN power devices technology was given. The main focus is put on transistors, i.e., 4H-SiC MOSFETs and GaN HEMTs. In particular, the importance of the channel mobility has been highlighted for both kind of devices. In 4H-SiC MOSFETs the most common trends to improve the channel mobility reported in literature are presented. Nitridations (NO or $N_2O$) remain the best processes to increase the channel mobility, without excessively compromising the device reliability.

In the case of GaN, the recessed hybrid MISHEMT is a currently debated solution to achieve a normally-off HEMT operation. For this technology, the choice of the dielectric and the control of its interface to GaN is fundamental to optimize the channel mobility and avoid a penalization of the $R_{ON}$ and of the $V_{th}$ stability..

**Acknowledgments.** The authors would like to thank the co-workers at CNR-IMM (F. Giannazzo, R. Lo Nigro, S. Di Franco, C. Bongiorno) for fruitful discussion and technical assistance. Colleagues of STMicroelectronics (F. Iucolano, A. Severino, S. Reina, A. Parisi, M. Saggio, S. Rascunà) are greatly acknowledged for support in device processing and characterization.

This work was partially supported by the ECSEL JU project WInSiC4AP (Wide Band Gap Innovative SiC for Advanced Power), Grant Agreement n. 737483.